\documentclass{sig-alternate}
\usepackage{mathptmx} 

\usepackage{fancyhdr}
\usepackage[normalem]{ulem}
\usepackage[hyphens]{url}
\usepackage[sort,nocompress]{cite}
\usepackage[final]{microtype}
\usepackage[keeplastbox]{flushend}
\usepackage{listings}
\usepackage{authblk}
\usepackage[bookmarks=true,breaklinks=true,letterpaper=true,colorlinks,linkcolor=black,citecolor=blue,urlcolor=black]{hyperref}

\fancypagestyle{firstpage}{
  \fancyhf{}
  
  \fancyhead[C]{White Paper v0.1}
  \fancyfoot[C]{\thepage}
}

\title{Asynchronous Memory Access Unit for General Purpose Processors} 
\author[1]{Luming Wang}
\author[1]{Xu Zhang}
\author[1]{Tianyue Lu}
\author[1]{Mingyu Chen}
\affil[1]{Institute of Computing Technology, Chinese Academy of Sciences, Beijing, China}

\begin{document}
\maketitle
\thispagestyle{firstpage}
\pagestyle{plain}


\begin{abstract}
  In future data centers, applications will make heavy use of far memory (including disaggregated memory pools and NVM). The access latency of far memory is more widely distributed than that of local memory accesses. This makes the efficiency of traditional blocking load/store in most general-purpose processors decrease in this scenario. Therefore, this work proposes an in-core asynchronous memory access unit.
  

\end{abstract}

\section{Introduction}

In recent years, more and more technologies aimed at improving the utilization of resources in cloud data centers have been proposed. More and more resources are organized into resources pool. Memory resources may be the next resources organized as a pool. Memory resources pool is similar to an old idea: distributed shared memory. However, the application of this idea at that time was limited due to the huge gap between network and memory in delay and bandwidth. At present, with the rapid development of network technology, these ideas are becoming increasingly attractive. However, nowadays the remote memory systems usually provide software interfaces (such as key-value, RDMA, files, etc) rather than load/store\cite{AguileraSoCC17}. 

Recently, new interconnect technologies and protocols (such as OpenCAPI, Gen-Z, CXL, etc) enable the construction of load/store interface disaggregated memory pool that contains multiple nodes. Prototypes of such systems have already been constructed by researchers\cite{Pinto2021ThymesisFlow}. It is foreseeable that complex disaggregated memory pools using load/store interface may emerge. However, the load/store interfaces are still not efficient enough.

On the other hand, Non-volatile Main Memory(NVMM) is starting to emerge, offering higher memory density and lower standby power consumption. However, it faces similar challenges as remote memory systems. Compared to traditional DRAM, NVMM has high latency and a wide range of latency variation (6x-30x higher write latency and 5x-10x higher read latency)\cite{liu_survey_2021}. Currently, there is not an efficient access interface for accessing NVM.  Commercial products, such as Intel's Optane DC, still provide a synchronous load/store interface. There is no mechanism provided for reacting to memory latency variation.

There are two main differences in accessing far memory compared to accessing traditional local memory.
\begin{description}
  \item[Widely distributed latency] The memory allocated from a disaggregate memory pool may locate on some faraway remote nodes. Furthermore, applications may use memory from DRAM, NVM, or other emerging memory devices. As a result, access latency becomes uncertain. Latency may distribute over a wide range.
  \item [Potential large aggregated bandwidth] As memory may come from multiple different machines, the aggregated bandwidth can increase significantly compared to local memory, making it a challenge of how to make use of the abundant bandwidth.
\end{description}

\begin{figure}[t]
\includegraphics[width=8cm]{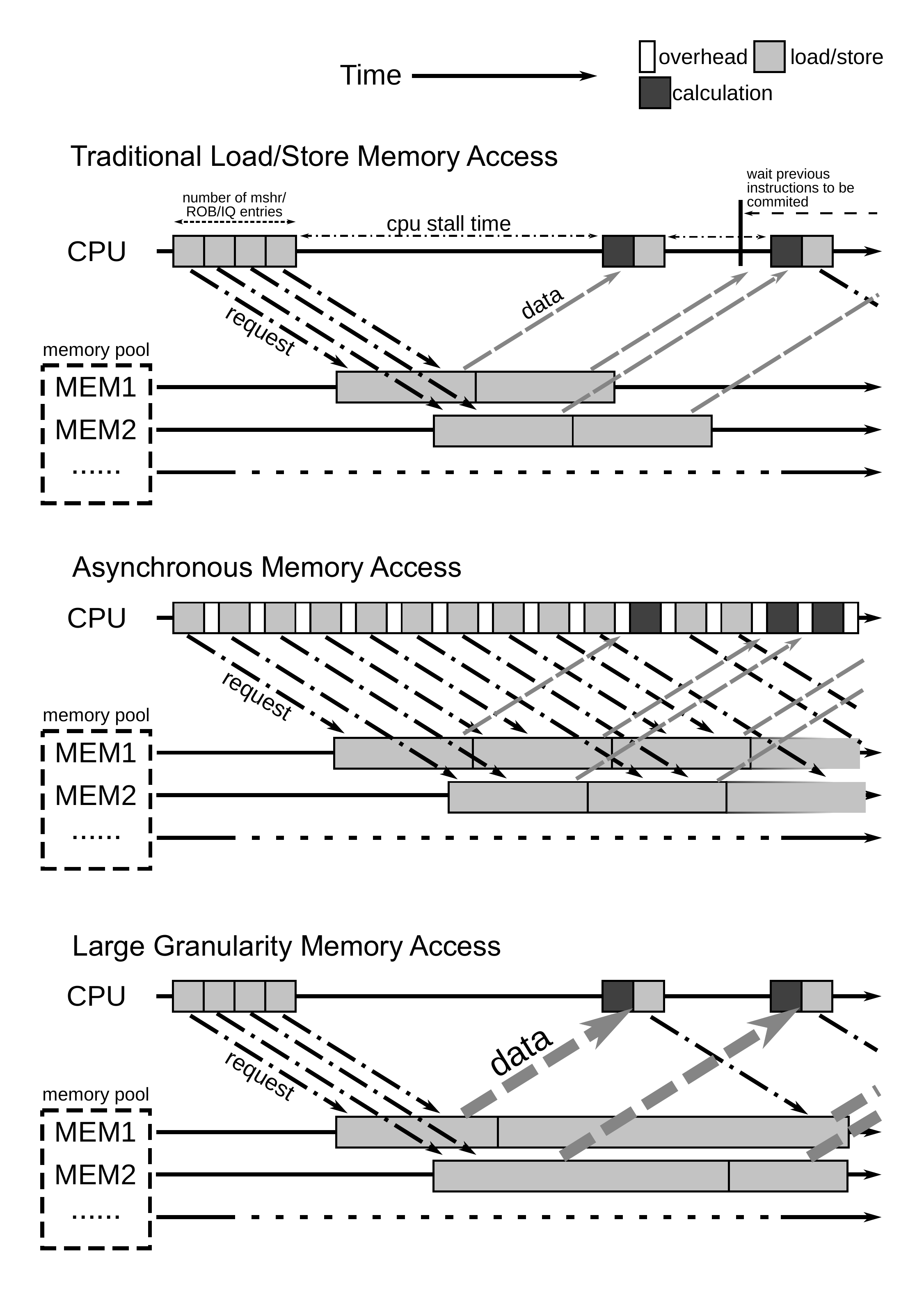}
\centering
\caption{Ways to improve memory bandwidth utilization\label{fig:motivation-timeline}}
\end{figure}

Access latency in traditional memory systems is also uncertain because of the multi-level cache hierarchy. However, the distribution of latency is relatively limited. The latency of a single access memory request is around 1ns (when L1 hits) to 200ns (when accesses local DRAM). Modern processors can tolerate this difference in latency by out-of-order execution and non-blocking cache. Fig \ref{fig:motivation-timeline} shows the limitations of current Out-of-Order processors in far memory scenarios. The range of latency they can tolerate is restricted by the number of entries in the instruction queue, ROB, and MSHRs. Once one of these resources is exhausted, the OoO processor cannot issue any more memory access requests. Moreover, once a long latency memory access instruction locates at the head of ROB, critical resources(such as ROB, IQ, etc.) will be occupied by it for a long time, which will cause degradation of the performance. Thus, it is difficult for modern processors to cope with the latency fluctuations (300ns-10us) caused by accessing far memory.

Although improving the out-of-order execution capability of traditional general-purpose processor cores (e.g., increasing the number of entries of MSHRs and ROB, using multi-level MSHRs and ROB, etc.) can also improve the performance of load/store in this scenario. But from such an improvement, even if it is feasible, it requires significant hardware resources. Another way is to use prefetching and multi-threading/coroutines techniques to hide access latency. However, the effectiveness of this approach is limited by the unavailability of observing access latency. In summary, the existing techniques have limited performance improvements for accessing far memory.

One approach to address this problem is asynchronous memory accessing. A similar predicament has already existed in network programming, where applications call blocking socket interfaces made program performance suffer from network latency. To solve this problem, asynchronous non-blocking interfaces such as select() and epoll() have been built. Just as network programming has evolved from the early synchronous blocking model to today's asynchronous non-blocking model, we believe that an asynchronous non-blocking model for memory access is also needed. Therefore, there should be some mechanisms for invoking asynchronous memory access.

Another approach is supporting memory accesses with variable granularity. As shown in Fig \ref{fig:motivation-timeline}, to improve performance, applications can initiate memory requests with a large granularity to fully utilize the bandwidth. Furthermore, applications can adjust the granularities based on data semantics to access memory flexibly and efficiently.

Concerning the above approaches, we propose an in-core Asynchronous Memory access Unit(AMU). The unit enables applications to asynchronously initiate complex variable granularity memory access requests by simple instructions. AMU can accomplish the extended memory access function with new asynchronous load/store instructions. Processors can still use original synchronous load/store instructions for compatibility while the data from both paths can transparently be consumed by computation instructions. We believe this is a more practical and efficient way than designing an un-core or off-chip accelerator. This white paper outlines the main features of the asynchronous memory access unit. 

\section{Asynchronous Memory Access Unit}
Asynchronous Memory access Unit is inspired by the Vector Processing Unit(VPU) of modern processors. The VPU is a separate functional unit in the CPU. Applications use the VPU through a standalone instruction set(i.e., vector instruction set), which contains a set of extra registers(i.e., vector registers) to hold the wide data to be processed. Besides, vector instructions are scheduled together with scalar instructions. Vector registers and scalar registers can exchange data efficiently.

Similarly, several asynchronous memory access instructions are designed for using AMU. The instruction that invokes a request can immediately be committed, once AMU receives the request. Thus, applications can continue to execute other operations rather than wait for memory access instructions to finish. Then, applications can poll whether there is a completed request.

Furthermore, each processor core is equipped with a ScratchPad Memory(SPM), which acts like vector registers in a vector instruction set. The reason for using SPM instead of register files is that the capacity and granularity of register files are limited. Therefore, SPMs are needed to hold the data for asynchronous memory accesses.

Data is moved asynchronously between SPM and memory by AMU. To initiate asynchronous memory access requests, applications can write several configuration registers, executing asynchronous memory access instructions. After receiving the request, AMU will move the data between memory and SPM in background.

For an application, the SPM is private memory space. Applications can use load/store instructions to load the data in the SPM into registers and process them with regular instructions. In addition, applications can copy data from local memory to the SPM and vice versa. The SPM is fully compatible with the processor's original data access and processing mechanisms.

\subsection{Instructions}
There are three core instructions of AMU. These instructions enable the most basic asynchronous memory access.

\begin{description}
  \item[Asynchronous load/store instructions] In AMU, \textit{aload/astore} instruction invokes a data movement request between SPM and memory. These two instructions have three operands(\textit{aload/astore Rd, Rs1, Rs2}). Operand \textit{Rs1} represents an SPM address. \textit{Rs2} represents a memory address. Memory accessing accelerator will move data between the provided SPM address and memory address. The request's id is stored into \textit{Rd}.
  \item[Instruction for getting a finished request's id] We propose \textit{getfin} for getting a completed request's id. If there is no finished request, the instruction returns a failure code. This design does not block execution regardless of whether there is a completed request or not.
\end{description}

\subsection{Registers}
Due to the limited length of instructions, some complex access memory settings cannot be encoded in the instructions. To solve this problem, we designed several configuration registers, which contain advanced configurations.

\begin{description}
  \item[Memory Access Configuration Register] This register contains advanced memory access configurations, including granularity, QoS labels, etc.
  \item[Default Configuration Register] Due to the limited encoding space of some instructions, it is not possible to encode all configurations. In particular, many asynchronous memory access instructions need to specify a memory access configuration register. For the case where the corresponding memory access configuration register cannot be specified in the instruction, the system automatically selects the configuration register specified in this register.
  \item[Access Pattern Register] The access pattern register is used to initiate complex asynchronous memory access. It contains the access pattern(such as stride, stream, etc.) of a complex memory access request.
\end{description}

In addition, some registers can be used to store software-defined configuration information. This is because, for a message interface based memory system\cite{chen_mims2014}, some software-defined information can be encoded in the message. These registers can increase the flexibility of the software.

\subsection{Programming Model}
Listing \ref{list:amae-basic} shows a basic example of asynchronous memory accessing. The code initiates an asynchronous memory access request with the aload instruction. The code then keeps trying to execute the getfin instruction to get the id of a completed request and can do other work while the request is pending. After the request completes, the code then reads the data from the SPM with the load instruction.
\begin{lstlisting}[language=c, caption=Asynchronous Memory Access Basic Example\label{list:amae-basic}]
  int memory_need_to_be_accessed;
  int *spm_space = (int*) A_SPM_ADDR;
  // Using aload to invoke an 
  // asynchronous memory access
  // requests. The request's id
  // is ignored here.
  aload(spm_space,
        &memory_need_to_be_accessed);
  while ((rd = getfin()) != 0) {
    // Do something else before
    // the request is finished.
  }
  // Access data from SPM via standard
  // load/store instruction
  printf("%d\n", *spm_space);
\end{lstlisting}

AMU instructions can support a variety of programming paradigms.
\subsubsection{Vector Model}
Vector instructions and vector processors are very early techniques for exploiting data-level parallelism. The SIMD instruction set, which is commonly supported in modern processors, uses a similar idea. As a technique to improve data-level parallelism, AMU instructions have many similarities in design ideas to vector instructions. Thus, it is possible to use AMU instructions for vector processing scenarios.

\subsubsection{Event-Driven Model}
The event-driven model is a common paradigm in single-thread non-blocking network programming. Furthermore, the \textit{aload}/\textit{astore} instructions are like non-blocking socket \textit{read()}/\textit{write()}. \textit{getfin} instructions are like the \textit{select()} in network programming. Thus, the event-driven model can be naturally applied to asynchronous memory accesses.

\subsubsection{Coroutin model}
For asynchronous access requests with complex access patterns, coroutines are a more suitable programming paradigm. Coroutines can easily work with high-performance concurrent data structures.

\section{Architecture Design}
\begin{figure}[t]
\includegraphics[width=8cm]{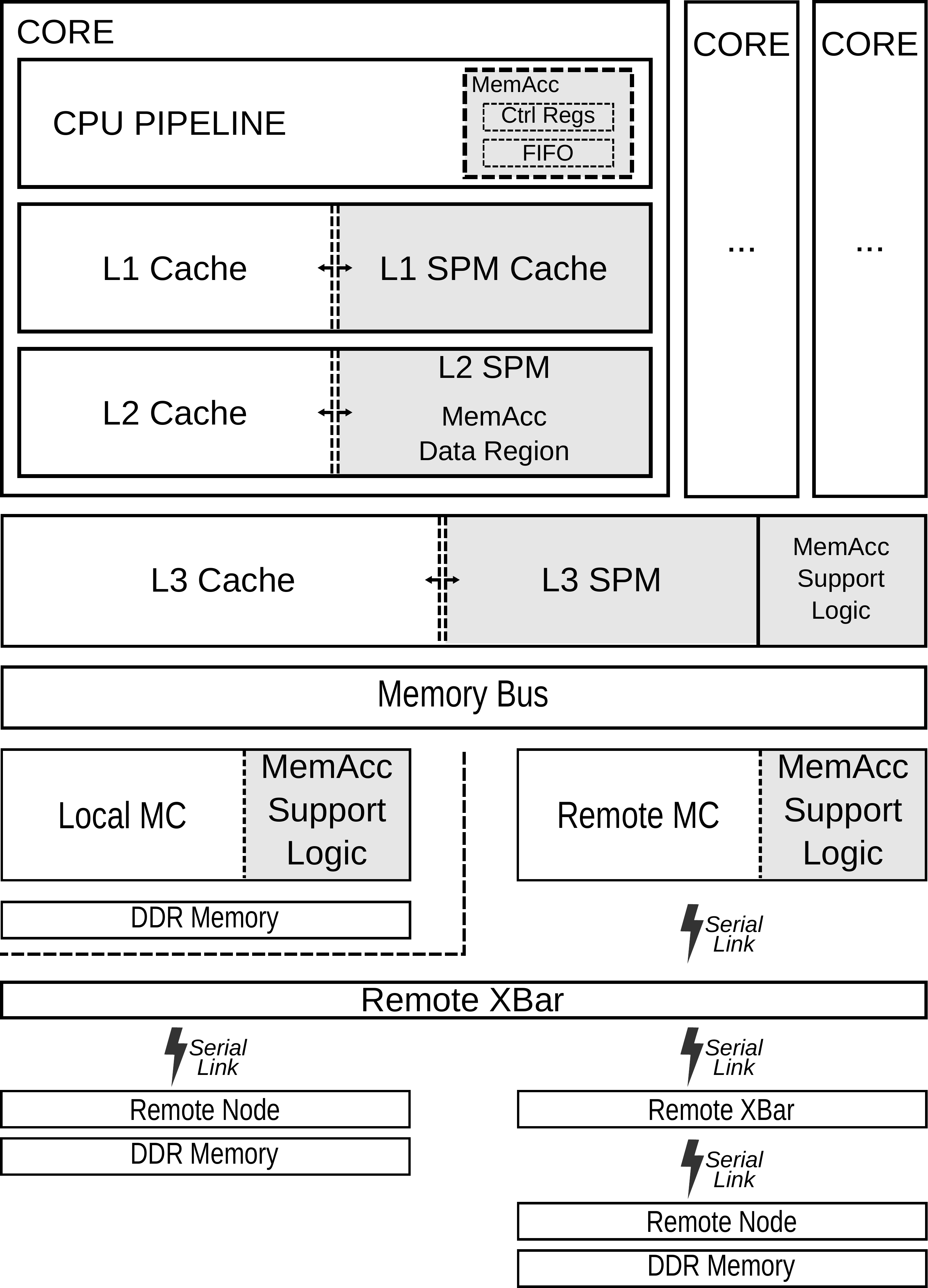}
\centering
\caption{Architecture Design\label{fig:arch-sim-overview}}
\end{figure}

Fig \ref{fig:arch-sim-overview} shows the architecture design of AMU. There are three key features of the architecture design:
\begin{description}
  \item[CPU Pipeline Integration] To support complex memory accessing scenarios, many configuration registers are integrated into CPU core's pipeline. Some of these registers indicate the accelerator's status, which allows programmers to rapidly gather the status of the accelerator. In addition, for the performance of asynchronous memory access instructions, the pipeline of the processor core needs to be modified to support the speculative execution of these instructions.
  \item[Reconfigurable Cache/SPM Space] AMU provides interfaces for applications to dynamically configure a part of the Cache as SPM. This design allows more flexibility for the software to decide how to use the accelerator. Applications can adjust the size of Cache and SPM themselves based on the workload. 
  \item[Integrated with L2 Controller] AMU's logic is integrated with the L2 controller. This logic implements the management and execution of asynchronous access requests and the movement between SPM and far memory.
\end{description}

\section{Future work}
In this paper, only the basic instructions and structures are presented. However, this design can easily support more memory extensions. For example, we can add some configuration registers and instructions for issuing processing-in-memory related requests. In addition, it can combine with message-based interface memory systems\cite{chen_mims2014}. It is possible to provide more configuration registers to issue memory access requests with richer semantics.




\bibliographystyle{IEEEtranS}
\bibliography{refs}

\begin{thebibliography}{1}
\providecommand{\url}[1]{#1}
\csname url@samestyle\endcsname
\providecommand{\newblock}{\relax}
\providecommand{\bibinfo}[2]{#2}
\providecommand{\BIBentrySTDinterwordspacing}{\spaceskip=0pt\relax}
\providecommand{\BIBentryALTinterwordstretchfactor}{4}
\providecommand{\BIBentryALTinterwordspacing}{\spaceskip=\fontdimen2\font plus
\BIBentryALTinterwordstretchfactor\fontdimen3\font minus
  \fontdimen4\font\relax}
\providecommand{\BIBforeignlanguage}[2]{{%
\expandafter\ifx\csname l@#1\endcsname\relax
\typeout{** WARNING: IEEEtranS.bst: No hyphenation pattern has been}%
\typeout{** loaded for the language `#1'. Using the pattern for}%
\typeout{** the default language instead.}%
\else
\language=\csname l@#1\endcsname
\fi
#2}}
\providecommand{\BIBdecl}{\relax}
\BIBdecl

\bibitem{AguileraSoCC17}
\BIBentryALTinterwordspacing
M.~K. Aguilera, N.~Amit, I.~Calciu, X.~Deguillard, J.~Gandhi, P.~Subrahmanyam,
  L.~Suresh, K.~Tati, R.~Venkatasubramanian, and M.~Wei, ``Remote memory in the
  age of fast networks,'' in \emph{Proceedings of the 2017 Symposium on Cloud
  Computing}, ser. SoCC '17.\hskip 1em plus 0.5em minus 0.4em\relax New York,
  NY, USA: Association for Computing Machinery, 2017, p. 121–127. [Online].
  Available: \url{https://doi.org/10.1145/3127479.3131612}
\BIBentrySTDinterwordspacing

\bibitem{chen_mims2014}
\BIBentryALTinterwordspacing
L.-C. Chen, M.-Y. Chen, Y.~Ruan, Y.-B. Huang, Z.-H. Cui, T.-Y. Lu, and Y.-G.
  Bao, ``\BIBforeignlanguage{en}{{MIMS}: {Towards} a {Message} {Interface}
  {Based} {Memory} {System}},'' \emph{\BIBforeignlanguage{en}{Journal of
  Computer Science and Technology}}, vol.~29, no.~2, pp. 255--272, Mar. 2014.
  [Online]. Available: \url{https://doi.org/10.1007/s11390-014-1428-7}
\BIBentrySTDinterwordspacing

\bibitem{liu_survey_2021}
\BIBentryALTinterwordspacing
H.-K. Liu, D.~Chen, H.~Jin, X.-F. Liao, B.~He, K.~Hu, and Y.~Zhang,
  ``\BIBforeignlanguage{en}{A {Survey} of {Non}-{Volatile} {Main} {Memory}
  {Technologies}: {State}-of-the-{Arts}, {Practices}, and {Future}
  {Directions}},'' \emph{\BIBforeignlanguage{en}{Journal of Computer Science
  and Technology}}, vol.~36, no.~1, pp. 4--32, Jan. 2021. [Online]. Available:
  \url{https://doi.org/10.1007/s11390-020-0780-z}
\BIBentrySTDinterwordspacing

\bibitem{Pinto2021ThymesisFlow}
C.~Pinto, D.~Syrivelis, M.~Gazzetti, P.~Koutsovasilis, A.~Reale, K.~Katrinis,
  and H.~P. Hofstee, ``Thymesisflow: A software-defined, hw/sw co-designed
  interconnect stack for rack-scale memory disaggregation,'' in \emph{2020 53rd
  Annual IEEE/ACM International Symposium on Microarchitecture (MICRO)}, 2020,
  pp. 868--880.

\end{thebibliography}

\end{document}